\documentstyle[preprint,prb,aps]{revtex}
\begin{document}
\draft
\title{Conductivity sum rule; comparison of coherent and incoherent
c-axis coupling}
\author{Wonkee Kim and J. P. Carbotte}
\address{Department of Physics and Astronomy, McMaster University, 
Hamilto, Ontario, Canada L8S 4M1}
\maketitle
\begin{abstract}
We calculate the $c$-axis kinetic energy difference between normal and 
superconducting state for coherent and for incoherent interlayer
coupling between CuO$_{2}$ planes. For coherent coupling the
ratio of the missing conductivity spectral weight to the superfluid
density is equal to one and there is no violation of the conventional
sum rule, but for the incoherent case we find it is always greater than
one whatever the nature of the impurity potential may be. To model more
explicitly 
YBa$_{2}$Cu$_{3}$O$_{7-x}$ around optimum doping, which is found to obey
the sum rule, we consider a plane-chain model and show that the sum rule
still applies. A violation of the sum rule of either sign is found
even for coherent coupling when the in-plane density of electronic states
depends on energy on a scale of the order of the gap.
\end{abstract}

It has been proposed that the interlayer coupling along the $c$ axis
of a high transition temperature $(T_{c})$ superconductor is incoherent, 
and the 
electronic kinetic energy along the $c$ axis changes when the system
enters the superconducting state.\cite{anderson,ekim,chakravarty,hirsh}
Recently, Basov {\it et al.}\cite{basov} have reported that there is a 
significant discrepancy between the superfluid density $\rho_{s}$
and the spectral weight missing from the real part of the $c$-axis
conductivity $N_{n}-N_{s}=
8\int^{\omega_{c}}_{0^{+}}d\omega\Bigl[\sigma^{n}_{1c}(\omega)
-\sigma^{s}_{1c}(\omega)\Bigr]$, where $\omega_{c}$ is a cutoff frequency
of the order of a bandwidth, 
in several high-$T_{c}$ cuprate
superconductors such as optimally doped Tl$_{2}$Ba$_{2}$Cu$_{6+x}$ (Tl$2201$).
This implies that the conventional sum rule
of Ferrel, Grover and Tinkham (FGT)\cite{fgt}
is violated. However,
the spectral discrepancy becomes vanishing
for YBa$_{2}$Cu$_{3}$O$_{6.85}$ and
disappears for the optimally doped
YBa$_{2}$Cu$_{3}$O$_{6.95}$ (YBCO) crystal with $T_{c}\simeq 93K$
as the $c$-axis response become more coherent with increasing 
oxygen content.\cite{basov,basov2} Basov {\it et al.}
also pointed out that
there is no such discrepancy in the in-plane response for any cuprate.
Moreover, for over-doped Tl$2201$, 
the sum rule discrepancy vanishes and a Drude-like peak develops in the 
conductivity for $T>T_{c}$.\cite{katz}
Theses observations, therefore, suggest that for coherent interlayer coupling
in the cuprate superconductors the conventional sum rule is obeyed.

In this paper, we consider both coherent and incoherent
$c$-axis coupling between CuO$_{2}$ planes.
For the coherent case we find that 
the superfluid density remains equal to
the missing optical spectral weight; in other words, it does not violate
the FGT sum rule. The $c$-axis kinetic energies in
the normal and superconducting state have the same value.
For incoherent $c$-axis coupling the ratio of the missing area to the
superfluid density is always larger than one in disagreement with
some recent experiments. In YBCO the CuO chains play an important role in the
electrodynamics and at the optimal doping a plane-chain model
\cite{atkinson,wu} is needed
to be complete.
Here we use this model
to investigate the $c$-axis conductivity sum rule. 
An algebraic calculation of the electronic kinetic energies is 
complicated and a numerical calculation is required
although it can be reduced a lot in a special case, in which
only the leading order in perturbation theory is kept.
This case is particularly interesting because it has been shown to exhibit
a pseudogap in the $c$-axis conductivity.\cite{wu}
Finally, we discuss the possibility that the FGT sum rule is violated 
in the plane-plane case when the in-plane density of states depends on
energy even for coherent case.

The Hamiltonian $H$ for a cuprate superconductor
with coherent $c$-axis coupling is $H=H_{0}+H_{c}$,
where $H_{0}$ describes a $d$-wave superconductor in a plane and
$H_{c}=\sum_{i\sigma}t_{\perp}\bigl[c^{+}_{i1\sigma}c_{i2\sigma}+
c^{+}_{i2\sigma}c_{i1\sigma}\bigr]$
is a coherent interlayer coupling due to the 
overlap of electronic
wave functions which is represented by $t_{\perp}$; therefore, 
by coherent coupling we mean
a tight binding like coupling along the $c$ axis.
It will not be necessary to treat $t_{\perp}$ as a constant in what follows.
It can depend on an angle in the plane.
For incoherent coupling the Hamiltonian is
$H'_{c}=\sum_{i\sigma}V_{i}\bigl[c^{+}_{i1\sigma}c_{i2\sigma}+
c^{+}_{i2\sigma}c_{i1\sigma}\bigr]$, where $V_{i}$ is an impurity scattering
potential, so that impurity scattering mediates the $c$-axis 
hopping and an impurity average is implied.\cite{hirschfeld,radtke} 

In the presence of an external vector potential $A_{z}$,
$H_{c}$ is modified to $H_{c}(A_{z})$ by the phase factor $\exp(-ieA_{z})$ for
$c^{+}_{i1\sigma}c_{i2\sigma}$ and $\exp(ieA_{z})$ for
$c^{+}_{i2\sigma}c_{i1\sigma}$.
For the response to an external field,
$H_{c}(A_{z})$ is expanded up to second order of $A_{z}$ to
obtain the current $j_{c}=-\delta H_{c}(A_{z})/\delta A_{z}
=j_{p}+j_{d}$, where
$j_{p}=-ied\sum_{i\sigma}t_{\perp}\bigl[c^{+}_{i1\sigma}c_{i2\sigma}-
c^{+}_{i2\sigma}c_{i1\sigma}\bigr]$ and $j_{d}=e^{2}d^{2}H_{c}A_{z}$
with $d$ the interlayer spacing.
In linear response theory, $\langle j_{c}\rangle=
\bigl[-\Pi+e^{2}d^{2}\langle H_{c}\rangle\bigr]A_{z}$,
where $\Pi$ is
the current-current correlation function associated with $j_{p}$
and $\langle H_{c}\rangle$ is the perturbation of $j_{d}$
due to $H_{c}$.
The conductivity $\sigma_{c}({\bf q},\omega)$ is given by
\begin{equation}
\sigma_{c}({\bf q},\omega)={i\over\omega}\Bigl[
\Pi({\bf q},\omega)-e^{2}d^{2}\langle H_{c}\rangle\Bigr].
\end{equation}
In the Matsubara formalism,
\begin{equation}
\Pi({\bf q},\omega)=2(ed)^{2}
T\sum_{\omega'}\sum_{\bf k}
t^{2}_{\perp}{\mbox T}{\mbox r}
\Bigl[{\hat\tau_{0}}{\hat G}({\bf k},\omega')
{\hat\tau_{0}}{\hat G}({\bf k},\omega'+\omega)
\Bigr],
\end{equation}
and
\begin{equation}
\langle H_{c}\rangle=2
T\sum_{\omega}\sum_{\bf k}
t^{2}_{\perp}{\mbox T}{\mbox r}
\Bigl[{\hat\tau_{3}}{\hat G}({\bf k},\omega){\hat\tau_{3}}
{\hat G}({\bf k},\omega)\Bigr],
\end{equation}
where 
${\hat\tau_{i}}$ is
the Pauli matrix in the spin space, and
${\hat G}({\bf k},\omega)$ is the Green's function in Nambu representation,
namely,
${\hat G}({\bf k},\omega)=-(i\omega{\hat\tau_{0}}+\xi_{\bf k}{\hat\tau_{3}}
-\Delta_{\bf k}\tau_{1})/(\omega^{2}+\xi^{2}_{\bf k}
+\Delta^{2}_{\bf k})$ wiht $\xi_{\bf k}$ the in-plane energy and
$\Delta_{\bf k}$ the gap which has $d_{x^2-y^2}$ symmetry in the cuprates.  

The $c$-axis conductivity sum rule\cite{hirsh,klein,chakravarty2}
of the system is
\begin{equation}
{2\over\pi}\int^{\infty}_{0}{}d\omega\sigma_{1c}(\omega)=
-e^{2}d^{2}\langle H_{c}\rangle.
\end{equation}
We use the unit such that $\hbar=c=k_{B}=1$ and set the volume of the system
to be unity.
From the sum rule, the superfluid density $\rho_{s}$ can be written as
\begin{equation}
\rho_{s}=8\int^{\omega_{c}}_{0^{+}}{}d\omega\Bigl[\sigma^{n}_{1c}(\omega)
-\sigma^{s}_{1c}(\omega)\Bigr]-4\pi e^{2}d^{2}
\Bigl[\langle H_{c}\rangle^{s}-\langle H_{c}\rangle^{n}\Bigr],
\label{rho}
\end{equation}
where
$\omega_{c}$ is the cutoff frequency for interband transitions that $H_{c}$
does not account for.

Since the difference between the superfluid density and
the missing spectral weight is proportional to the kinetic energy
difference between normal and superconducting state as seen in 
Eq. (\ref{rho}),
it is necessary to calculate 
$\langle H_{c}\rangle^{s}-\langle H_{c}\rangle^{n}$ to see if
the FGT sum rule is violated by coherent $c$-axis coupling.
For the normal state,
\begin{equation}
\langle H_{c}\rangle^{n}=
4T\sum_{\omega}\sum_{\bf k}
t^{2}_{\perp}G_{0}({\bf k},\omega)^{2},
\end{equation}
where 
$G_{0}({\bf k},\omega)$ is a normal state Green's function
and $t_{\perp}$ may depend on $k_{z}$ and $\phi=\tan^{-1}(k_{y}/k_{x})$.
We assume a cylindrical Fermi surface with $\xi=k^{2}/2m-\mu$,
where $\mu$ is a chemical potential in the plane,
and a $d$-wave gap $\Delta_{{\bf k}}=\Delta(T)\cos 2\phi_{k}$.
Then, we obtain
\begin{eqnarray} 
\langle H_{c}\rangle^{n}&&=4\sum_{k_{z}}
\int{}{d\phi\over2\pi}t^{2}_{\perp}\int^{\omega_{c}}_{-\omega_{c}}
d\xi N(\xi){\partial f(\xi)\over\partial\xi}
\nonumber\\
&&=-4N(0)\sum_{k_{z}}
\int{}{d\phi\over2\pi}t^{2}_{\perp}
\tanh({\omega_{c}\over 2T}),
\label{nkin}
\end{eqnarray}
where the integration range is limited by $\omega_{c}$, and 
the density of states, $N(\xi)$, is approximated by a constant value 
$N(0)$ around the Fermi energy.
Later, we will discuss the effect of $N(\xi)$ on $\langle H_{c}\rangle$
and will note
the possibility that the FGT sum rule may be violated even for
coherent $c$-axis coupling.
Since ${\partial f(\xi)/\partial\xi}=-\delta(\xi)$ at zero temperature ($T=0$), 
$\langle H_{c}\rangle^{n}$ turns out to be 
$-4N(0)\sum_{k_{z}}\int{}d\phi/(2\pi)t^{2}_{\perp}$.
For a superconducting state with superconducting Green's functions
$G({\bf k},\omega)$ and $F({\bf k},\omega)$,
\begin{eqnarray}
\langle H_{c}\rangle^{s}&&=
4T\sum_{\omega}\sum_{\bf k}
t^{2}_{\perp}\Bigl[G({\bf k},\omega)^{2}-F({\bf k},\omega)^{2}\Bigr]
\nonumber\\
&&=-4T\sum_{\omega}\sum_{k_{z}}\int{}{d\phi\over2\pi}t^{2}_{\perp}
\int^{\omega_{c}}_{-\omega_{c}}
d\xi N(\xi){{\omega^{2}-\xi^{2}+\Delta^{2}_{\bf k}}\over
{(\omega^{2}+\xi^{2}+\Delta^{2}_{\bf k})^{2}}}
\nonumber\\
&&=-4N(0)\sum_{k_{z}}\int{}{d\phi\over2\pi}t^{2}_{\perp}
{\omega_{c}\over\sqrt{\omega^{2}_{c}+\Delta^{2}_{\bf k}}}
\tanh({\sqrt{\omega^{2}_{c}+\Delta^{2}_{\bf k}}\over 2T}).
\label{skin}
\end{eqnarray}
The difference between $\langle H_{c}\rangle^{s}$
and $\langle H_{c}\rangle^{n}$ is of the order of 
$(\Delta(T)/\omega_{c})^{2}$; therefore,
coherent $c$-axis coupling does not violate the FGT sum rule
as long as $\omega_{c} >> \Delta(0)$
even if $t_{\perp}$ depends on $\phi$. Note that the difference is
largest at $T=0$ and vanishes as $T\rightarrow T_{c}$.

The calculations for incoherent (impurity mediated) $c$-axis coupling
proceed in the same way as before.
Note that in this case 
$j_{p}=-ied\sum_{i\sigma}V_{i}\bigl[c^{+}_{i1\sigma}c_{i2\sigma}-
c^{+}_{i2\sigma}c_{i1\sigma}\bigr]$ and $j_{d}=e^{2}d^{2}H'_{c}A_{z}$,
and an impurity configuration average is required.
We derive the normalized missing spectral weight 
$(N_{n}-N_{s})/\rho_{s}$ under assumption of
a constant density of states and show that it is greater than one.

The penetration depth $\lambda_{c}$ can be calculated in two ways.
Based on the Kramers-Kronig relation for the conductivity,
we obtain $\lambda_{c}$, namely,
$1/4\pi\lambda^{2}_{c}=\lim_{\omega\rightarrow 0}
[\omega\mbox{I}\mbox{m}\sigma_{c}(0,\omega)]$.
Alternatively, using Eq.(\ref{rho}) 
we can also calculate $\lambda_{c}(=1/\sqrt{\rho_{s}})$.
Equate these two expressions of $\lambda_{c}$,
then after integration over energy
we arrive the formula as follows:
\begin{equation}
{(N_{n}-N_{s})\over\rho_{s}}=
{1\over2}+
{1\over2}{\sum_{\omega}\int d\phi_{k} d\phi_{p}|V(\phi_{k},\phi_{p})|^2
\Bigl[1-{
\omega^2\over
\sqrt{\omega^2+\Delta^{2}_{k}}\sqrt{\omega^2+\Delta^{2}_{p}}}
\Bigr]
\over
\sum_{\omega}\int d\phi_{k} d\phi_{p}|V(\phi_{k},\phi_{p})|^2
{\Delta_{k}\over\sqrt{\omega^2+\Delta^{2}_{k}}}
{\Delta_{p}\over\sqrt{\omega^2+\Delta^{2}_{p}}}}.
\label{ratio}
\end{equation}
The second term in Eq. (\ref{ratio}) can easily be shown to be bigger than
one half whatever the angular dependence of the impurity potential
$V(\phi_{k},\phi_{p})$ may be.\cite{asymp} Thus the normalized missing
spectral weight is always greater than one. In a simple model of
impurity scattering,\cite{ekim,hirschfeld}
for which $|V(\phi_{k},\phi_{p})|^{2}=
|V_{0}|^{2}+|V_{1}|^{2}\cos 2\phi_{p}\cos 2\phi_{k}$,
we found that $(N_{n}-N_{s})/\rho_{s}\ge1.58$.
This incoherent coupling model, therefore, 
does not agree with recent findings. The sum rule
is one for YBCO around optimum doping indicating coherent $c$-axis
coupling and less than one for the underdoped case. To treat YBCO around 
optimum doping more realistically
we need to include the complications introduced
by existence of the chains along the $b$ axis. 

Penetration depth ($\lambda_{a(b)}$) experiments
\cite{zhang}
in YBCO have shown that both $\lambda_{a}$ and $\lambda_{b}$ are linear
$T$ at a low temperature and that 
a considerable amount of the condensate resides on the chains.
To treat this case we need to consider a plane-chain coupling model.
\cite{atkinson,wu}
We assume the hybridization of Fermi surfaces between plane
and chain arising through coherent coupling. For simplicity we also
assume the gap in the chain has a $d$-wave symmetry 
and its magnitude is of the order of that in the plane.
The Hamiltonian for a coupled plane-chain system is 
$H=\sum_{{\bf k}}{\hat C}^{+}_{{\bf k}}{\hat h}_{{\bf k}}{\hat C}_{{\bf k}}$,
where 
${\hat C}^{+}_{{\bf k}}=\Bigl(C^{+}_{1{\bf k}\uparrow}, C_{1-{\bf k}\downarrow},
C^{+}_{2{\bf k}\uparrow}, C_{2-{\bf k}\downarrow}\Bigr)$ and 
\begin{equation}
{\hat h}_{{\bf k}}=
        \left(\begin{array}{cccc}
         \xi_{1{\bf k}}&-\Delta_{1{\bf k}}&t(k_{z})&0\\
         -\Delta_{1{\bf k}}&-\xi_{1{\bf k}}&0&-t(k_{z})\\
         t(k_{z})&0&\xi_{2{\bf k}}&-\Delta_{2{\bf k}}\\
         0&-t(k_{z})&-\Delta_{2{\bf k}}&-\xi_{2{\bf k}}
\end{array}\right)
\end{equation}
where $t(k_{z})=-t_{0}\cos(k_{z}d/2)$ for coherent coupling
between plane and chain,
$\xi_{1(2)}$ is the energy dispersion in the plane (chain), and
$\Delta_{1(2)}$ is a gap of the plane (chain).
We point out here that
the conclusion we make later does not depends on the simple form
of $t(k_{z})$, and that $\Delta_{1(2)}$ and $\xi_{1(2)}$
depend only on $k_{x}$ and $k_{y}$. 

The Hamiltonian of the plane-chain coupling model is also decomposed
into two parts, $H=H_{0}+H_{c}$.
$H_{0}$ is for the superconductivity
in the plane-chain coupling system and its eigenvalues can be reduced to
$\pm E_{\pm}=\pm\sqrt{\epsilon^{2}_{\pm}+\Delta^{2}_{\bf k}}$, where
$\epsilon_{\pm}$ are normal state energy dispersions
$\epsilon_{\pm}={(\xi_{1}+\xi_{2})/2}\pm
\sqrt{(\xi_{1}-\xi_{2})^{2}/4+t(k_{z})^{2}}$
with $\Delta_{1\bf k}=\Delta_{2\bf k}=\Delta_{\bf k}$ for simplicity.
Extensive work on this Hamiltonian
can be found in Ref.\cite{atkinson}

In order to calculate the linear response of the system to the external
electromagnetic field, we modify $H_{c}$ with the phase factor 
mentioned before and follow the same procedure to derive
the current $j_{c}=j_{p}+j_{d}$. Then, we obtain
\begin{equation}
j_{p}={edt_{0}\over2}\sum_{{\bf k}}\sin(k_{z}d/2)
{\hat C}^{+}_{{\bf k}}
{\hat\sigma_{1}}\otimes{\hat\tau_{0}}{\hat C}_{{\bf k}},
\end{equation}
where $d/2$ is the distance between a plane and a chain and 
${\hat\sigma_{1}}$ is a Pauli matrix in the plane-chain space, and
$j_{d}=e^{2}(d/2)^{2}H_{c}A_{z}$, where 
$H_{c}=\sum_{{\bf k}}t(k_{z})
{\hat C}^{+}_{{\bf k}}
{\hat\sigma_{1}}\otimes{\hat\tau_{3}}{\hat C}_{{\bf k}}.$
The $c$-axis conductivity for ${\bf q}=0$, 
$\sigma_{c}(0,\omega)$, of the sysyem is also derived to be
$\sigma_{c}(0,\omega)=(i/\omega)\Bigl[
\Pi(0,\omega)-e^{2}(d/2)^{2}\langle H_{c}\rangle\Bigr]$,
where
\begin{equation}
\Pi(0,\omega)=(edt_{0}/2)^{2}T\sum_{\omega'}\sum_{\bf k}\sin^{2}(k_{z}d/2)
{\mbox T}{\mbox r}
\Bigl[{\hat\sigma_{1}}\otimes{\hat\tau_{0}}{\hat G}({\bf k},\omega')
{\hat\sigma_{1}}\otimes{\hat\tau_{0}}{\hat G}({\bf k},\omega'+\omega)
\Bigr],
\end{equation}
and
\begin{equation}
\langle H_{c}\rangle=t^{2}_{0}T\sum_{\omega}
\sum_{\bf k}\cos^{2}(k_{z}d/2)
{\mbox T}{\mbox r}
\Bigl[{\hat\sigma_{1}}\otimes{\hat\tau_{3}}{\hat G}({\bf k},\omega)
{\hat\sigma_{1}}\otimes{\hat\tau_{3}}
{\hat G}({\bf k},\omega)\Bigr],
\label{kinetic}
\end{equation}
with ${\hat G}({\bf k},\tau)=-\langle{\cal T}
[{\hat C}_{\bf k}(\tau){\hat C}^{+}_{\bf k}(0)]
\rangle$, which is a $(4\times4)$ matrix. The Green's function
${\hat G}({\bf k},\omega)$ is given by
${\hat G}({\bf k},\omega)=(i\omega-{\hat h}_{\bf k})^{-1}$. 
We emphasize that
the Hamiltonian in this model is quite different from the usual macroscopic
tunneling Hamiltonian,\cite{mahan} for which, for example,
${\hat G}_{13}({\bf k},\tau)=
-\langle{\cal T}[C_{1{\bf k}\uparrow}(\tau)
C^{+}_{2{\bf k}\uparrow}(0)]\rangle$ is not allowed because
each layer is independent (as is the case for the previous
coupling model);
however, it is possible in the present model 
because of the hybridization through
the chain between the two Fermi surfaces of plane and chain.
Introducing a unitary
matrix $U$ which diagonalizes ${\hat h}_{\bf k}$, one can show
${\hat G}_{ij}({\bf k},\omega)=
\sum^{4}_{m=1}U_{im}U^{+}_{mj}/(i\omega-E_{m})$, where 
$E_{m}=\pm E_{\pm}$ if $\Delta_{1\bf k}=\Delta_{2\bf k}$.

$\langle H_{c}\rangle$ becomes complicated and 
the energy dispersion in the chain is quite different from
that in the plane so that a numerical calculation is required
to see if the difference between 
$\langle H_{c}\rangle^{s}$ and $\langle H_{c}\rangle^{n}$
is negligible.
However, since $t_{0}$ in Eq. (\ref{kinetic}) is assumed small
we may expand $\langle H_{c}\rangle$
in terms of $t_{0}$ and keep only the leading order,
which is $t^{2}_{0}$.
This case includes only interband Hamiltonian but is still very interesting
as it can exhibit a $c$-axis pseudogap.\cite{wu}
In this approximation with $\Delta_{1\bf k}=\Delta_{2\bf k}
=\Delta_{\bf k}$ and for $\mu'=\mu$ as a special case, 
$\langle H_{c}\rangle^{s}$ becomes
\begin{equation}
\langle H_{c}\rangle^{s}=-4T\sum_{\omega}\sum_{\bf k}t(k_{z})^{2}
{{\omega^{2}-\xi_{1}\xi_{2}+\Delta^{2}_{\bf k}}\over
{(\omega^{2}+\xi^{2}_{1}+\Delta^{2}_{\bf k})
(\omega^{2}+\xi^{2}_{2}+\Delta^{2}_{\bf k})}}.
\label{pckin}
\end{equation}
Note that $\langle H_{c}\rangle^{s}$ in Eq. (\ref{pckin}) is almost same as
$\langle H_{c}\rangle^{s}$ in Eq. (\ref{skin}) for the simple coherent coupling
case except that now $\xi_{1}\ne\xi_{2}$ 
and $d/2$ appears rather than $d$ in $t(k_{z})$.
One, therefore, may expect that $\delta\langle H_{c}\rangle$ will
vanish to order $(\Delta(T)/\omega_{c})^{2}$. It is obvious that
$\delta\langle H_{c}\rangle$ is identically zero along the nodal
lines, and $\delta\langle H_{c}\rangle$
is largest along the anti-nodal directions.
Since $\xi_{1}=k^{2}/2m-\mu$ and $\xi_{2}=k^{2}_{y}/2m-\mu$, we
introduce $\xi=\xi_{1}$, then $\xi_{2}=\xi\sin(\phi)^{2}-\mu\cos(\phi)^{2}$.
If $\phi=\pi/2$, then $\xi_{2}=\xi$; therefore, it can be seen that
$\delta\langle H_{c}\rangle_{\phi=\pi/2}$ 
is of the order of $(\Delta/\omega_{c})^{2}$.
For $\phi=0$, $\xi_{2}=-\mu$ and it can be shown that
\begin{eqnarray}
\langle H_{c}\rangle^{s}_{\phi=0}\simeq &&2N(0)\sum_{k_{z}}t^{2}_{\perp}
\int^{\omega_{c}}_{-\omega_{c}}{}d\xi
\Bigl[ {\mu\over{\xi^{2}-\mu^{2}}}
\tanh({\mu\over2T})
\nonumber\\
&&-{\xi^{2}\over{(\xi^{2}-\mu^{2})\sqrt{\xi^{2}+\Delta^{2}}}}
\tanh({\sqrt{\xi^{2}+\Delta^{2}}\over 2T})\Bigr].
\end{eqnarray}
Now, the leading order of $\delta\langle H_{c}\rangle_{\phi=0}$
changes to $(\Delta/\mu)^{2}$.
It is possible to show that for an arbitrary $\phi$, 
as long as $\mu$ and $\omega_{c} >> \Delta$,
$\delta\langle H_{c}\rangle_{\phi}$ is negligible, and consequently,
the FGT sum rule is not violated in the plane-chain coupling model.

In a numerical calculation  for the general case 
without the above simplification, 
we have computed
$\delta\langle H_{c}\rangle/\langle H_{c}\rangle^{s}$, which is
the fractional change in kinetic energy. We have taken
$\xi_{2}=k^{2}_{y}/2m-\mu'$ for the chain energy dispersion, where
$\mu'$ is a chemical potential in the chain. For simplicity, we also assume
$\mu'=\mu$. It has been shown that $\mu'-\mu(<<\mu)$ may
correspond to the pseudogap seen in the $c$-axis response
of over-doped YBCO;\cite{wu} however,
it makes no difference in the numerical evaluation of
$\delta\langle H_{c}\rangle/\langle H_{c}\rangle^{s}$. We choose
$T=12K$, $\Delta(T)=20$meV, $t_{0}=2$meV, $\mu=500$ meV and
$\omega_{c}=400$meV.
We found that $\delta\langle H_{c}\rangle/\langle H_{c}\rangle^{s}$
becomes more negligible as we increase the summation range of
the Matsubara frequency $\omega$. 
For $|\omega|\le200\pi T$,
$\delta\langle H_{c}\rangle/\langle H_{c}\rangle^{s}\simeq 8.8\times10^{-3}$,
and for $|\omega|\le2000\pi T$,
$\delta\langle H_{c}\rangle/\langle H_{c}\rangle^{s}\simeq 5.4\times10^{-3}$.

One may consider a plane-plane coupling {\it through a chain}. In order
to investigate the $c$-axis kinetic energy for such a coupling,
one needs to replace $\xi_{2}$ and $\Delta_{2}$ with $\xi_{1}$ and $\Delta_{1}$,
respectively. For the hopping amplitude, $t(k_{z})$ can be simply
changed to $t(k_{z})^{2}$ because the plane-chain and chain-plane distances
are the same and equal to $d/2$. Then,
one can algebraically show 
that $\delta\langle H_{c}\rangle$ is as negligible 
as before. It is also possible to see that
$\delta\langle H_{c}\rangle$ has a symmetry with respect to
$\xi_{1}\leftrightarrow\xi_{2}$ and $\Delta_{1}\leftrightarrow\Delta_{2}$;
in other words, $\delta\langle H_{c}\rangle$ for the {\it chain-plane}
coupling is the same as that for the {\it plane-chain} coupling.
Therefore, it implies that
$\delta\langle H_{c}\rangle$ along the $c$ axis is conserved 
for coherent coupling.

So far we have taken the density of states as a constant: 
$N(\xi)=N(0)$ $(-\omega_{c}\le\xi\le\omega_{c})$ and concluded that
the difference of the $c$-axis electronic kinetic energies
between normal and supercondcuting state is negligible.
Now we would like to consider the
effect of $N(\xi)$ on the sum rule when it is a function of $\xi$
to illustrate possible changes. 
If it varies strongly with $\xi$, it clearly cannot be
approximated by $N(0)$. 
We taylor-expand $N(\xi)$ up to $\xi^{2}$ near $\xi=0$.
Then, $N(\xi)=N(0)+\xi\partial N(\xi)/\partial\xi|_{0}+(\xi/2)^{2}
{\partial^{2}N(\xi)/\partial\xi^{2}}|_{0}$. At $T=0$, 
$\langle H_{c}\rangle^{n}$ of Eq. (\ref{nkin}) does not change; however,
Eq. (\ref{skin}) for $\langle H_{c}\rangle^{s}$ changes due to 
$(\xi/2)^{2}N"(0)$, where $N"(0)=\partial^{2}N(\xi)/\partial\xi^{2}|_{0}$.
Assuming $t_{\perp}$ in Eqs. (\ref{nkin}) and 
(\ref{skin}) does not depend on $\phi$,
we obtain 
\begin{equation}
\delta\langle H_{c}\rangle/\langle H_{c}\rangle^{n}\simeq
(8N(0))^{-1}N"(0)\Delta(0)^{2}\ln(\omega_{c}/\Delta(0)).
\end{equation}
Note that this correction can have either sign depending on the 
sign of the second derivative. 
For $N"(0)/N(0)\sim 1/\omega^{2}_{c}$,
$\delta\langle H_{c}\rangle/\langle H_{c}\rangle^{n}\sim x^{-1.65}$, where
$x=\omega_{c}/\Delta(0)$,
because $\ln(x)/x^{2}\simeq x^{-1.65}$ when $x >> 1$. 
If $N"(0)/N(0)\sim 1/(\Delta(0)\omega_{c})$,
then $\delta\langle H_{c}\rangle/\langle H_{c}\rangle^{n}\sim x^{-0.65}$;
thus, $\delta\langle H_{c}\rangle$ is considerable. For this to be 
the case $N"(0)$ needs to exhibit variation on an energy scale of order
$\Delta$ rather than $\omega_{c}$. In a realistic model a Taylor expansion 
about $\xi=0$ may not be accurate but our calculations serve to illustrate
the main point. Violation of the FGT sum rule of either sign can result from
an energy dependence in the in-plane electronic density of states $N(\xi)$.
The exact amount depends on details and cannot be known without a specific
knowledge of the band structure involved. In-plane dynamics gets reflected
in $c$-axis properties.

For coherent interlayer coupling between CuO$_{2}$ planes the superfluid 
density is equal to the missing optical weight; the FGT sum rule is satisfied.
This applies even in more realistic model for YBCO around optimum
doping such as the plane-chain model with two atoms per unit cell.
On the other hand incoherent $c$-axis coupling mediated through impurity
scattering gives a sum rule which is always larger than one 
and is in disagreement
with experiment. 
To get the sum rule to be less than one as is observed in underdoped
YBCO and other systems such as optimally doped Tl2201, it may be necessary
to go to more exotic non-Fermi liquid pseudogap model for the in-plane
motion as discussed recently by Ioffe and Millis.\cite{millis} 
Their arguments however do not apply to optimally doped
Tl2201  because this system does not show a pseudogap. 
Their pseudogap argument that leads to the cancellation of 
$G({\bf k},\omega)$ and $G_{0}({\bf k},\omega)$ contribution to
the ratio of missing area to superfluid density making it one half
instead of one for the preformed pair model was made for coherent c-axis
coupling, but we find it also applies to the incoherent case.\cite{asymp}
Another interesting
model for the in-plane dynamics is the "mode" model of
Norman {\it et al.}\cite{norman} introduced from consideration of ARPES
data. In more conventional models a sum rule violation of either
sign can also be obtained
if there is a strong energy dependence to the density of states near the
Fermi surface on the scale of a few times the gap.

W.K. is grateful to N. D. Whelan and C. Kallin
for useful discussions and J.P.C. to D. Basov for discussions.
This work was supported in part by the Natural Sciences and Engineering
Research Council of Canada (NSERC) and by the Canadian Institute for 
Advanced Research (CIAR).


\begin{references}
\bibitem{anderson} P. W. Anderson, {\it The Theory of Superconductivity
in the High-$T_{c}$ Cuprates} (Princeton Univ. Press, Princeton,
NJ, 1997); P. W. Anderson, Science {\bf 279}, 1196 (1998).

\bibitem{ekim} E. H. Kim, Phys. Rev. B {\bf 58}, 2452 (1997).

\bibitem{chakravarty} S. Chakravarty, Eur. Phys. J. B {\bf 5},
337 (1998).

\bibitem{hirsh} J. E. Hirsh, Physica C {\bf 199}, 305 (1992);
J. E. Hirsh, Physica C {\bf 201}, 347 (1992).

\bibitem{basov} D. N. Basov {\it et al.},
Science {\bf 283}, 49 (1999).

\bibitem{fgt} R. A. Ferrel and R. E. Glover, Phys. Rev. {\bf 109},
1398 (1958); M. Tinkham and R. A. Ferrel, Phys. Rev. Lett. {\bf 2},
331 (1959).

\bibitem{basov2} D. N. Basov (unpublished).

\bibitem{katz} A. S. Katz {\it et al.},
cond-mat/9905170

\bibitem{atkinson} W. A. Atkinson and J. P. Carbotte,
Phys. Rev. B {\bf 52} 10 601 (1995); {\it ibid.}
{\bf 55}, 3230 (1997); {\it ibid.} {\bf 55}, 12 748 (1997).

\bibitem{wu} W. C. Wu, W. A. Atkinson and J. P. Carbotte,
J. Supercond. {\bf 11}, 305 (1998).

\bibitem{hirschfeld} P. J. Hirschfeld, S. M. Quinlan, and D. J. Scalapino,
Phys. Rev. B {\bf 55},12 742 (1997).

\bibitem{radtke} R. J. Radtke, V. N. Kostur, and K. Levin,
Phys. Rev. B {\bf 53}, R522 (1996).

\bibitem{klein} M. V. Klein and G. Blumberg, Science {\bf 283}, 42 (1999).

\bibitem{chakravarty2} S. Chakravarty, H. -Y. Kee, and E. Abrahams, 
Phys. Rev. Lett. {\bf 82}, 2366 (1999).

\bibitem{asymp} We assume the infinite band approximation
($\omega_{c}\rightarrow\infty$) for simplicity. 
However, it is generally true even for a finite $\omega_{c}$. Note
that $|V(\phi_k,\phi_p)|^2$ is positive definite. For more details
see W. Kim, N. D. Whelan and J. P. Carbotte (to be published elsewhere).

\bibitem{zhang} K. Zhang {\it et al.},
Phys. Rev. Lett. {\bf 73}, 2484 (1994);
D. N. Basov {\it et al.}, 
Phys. Rev. Lett. {\bf 74}, 598 (1995).

\bibitem{mahan} See, for example, G. D. Mahan, {\it Many-Particle Physics}
(Plenum, New York, 1990).

\bibitem{millis} L. B. Ioffe and A. J. Millis, Science {\bf 285}, 1241 (1999).

\bibitem{norman} M. R. Nornam {\it et al.}, cond-mat/9912043.

\end{references}
\end{document}